\documentclass[aps,showpacs,twocolumn,longbibliography,nofootinbib,secnumarabic]{revtex4-1}

\usepackage{graphicx}
\usepackage{float}
\usepackage{xcolor}
\usepackage{amsmath}
\begin{document}
\title
{\bf Entanglement criteria for two strongly interacting ensembles}
\author{Mehmet G\"{u}nay$^{\bf (1)}$} \email{gunaymehmt@gmail.com}
\author{\"{O}zg\"{u}r Esat M\"{u}stecapl\i o\u{g}lu$^{\bf (2)}$}
\author{Mehmet Emre Tasgin$^{\bf (1)}$}
\affiliation{ ${\bf (1)}$ { Institute  of  Nuclear  Sciences, Hacettepe University, 06800 Ankara, Turkey}
\break
${\bf (2)}$ {Department of Physics, Ko\c{c} University, \.{I}stanbul,  Sar{\i}yer, 34450, Turkey}}
\begin{abstract}
Two interacting atomic ensembles display a Dicke-like quantum phase transition above a critical coupling strength. We show that an ensemble-ensemble entanglement accompanies the quantum phase transition. We derive entanglement criteria, which can witness the entanglement of the two interacting ensembles. We observe that all criteria are successful in the thermodynamic limit, while only the newly introduced ones~(number squeezing-like criteria) can witness the ensemble-ensemble entanglement for a finite number of particles. We also mention about implementations of these criteria to two-component condensate systems and nanoscale quantum plasmonics.
\end{abstract}
\maketitle
\section{Introduction}

Quantum entanglement is the key resource at the heart of cutting edge quantum technologies, such as quantum teleportation systems for secure communication~\cite{QuantumTeleportation143km,BraunsteinNaturePhot2015,SatelliteTeleportation2017}, quantum memories for quantum computers~\cite{heshami2016quantum}, and atom clocks and interferometers for quantum sensing and metrology~\cite{degen2017quantum}. Further improvement of these technologies demand scaling up to the entangled systems and at the same time, minimization of decoherence. Advantages of many particle entanglement~(MPE) has been revealed particularly in fast, robust, and secure high capacity data storage~\cite{heshami2016quantum,wineland1992spin,huelga1997improvement,giovannetti2006quantum, ScullyScience2009SR,vernaz2018highly,ScullyPRL2015Subradiance}, metrology below standard quantum limit of spin noises~\cite{QuantMetrology_RevModPhys_2018} and improving power of quantum heat engines~\cite{hardal2015superradiant}.

Besides well-known spin squeezing-like interactions~\cite{sorensen2001Nature,SqueezingTransferSpin_PRL_1997,Kitagawa&UedaPRA1993,Tasgin&MeystrePRA2011,MeasurementInducedSqueezing_PRA_1999,kuzmich2000generation},  MPE can be generated also via inducing quantum phase transitions~(QPT), e.g. superradiant Dicke QPT.  Dicke model describes interaction of a single mode quantized field with an ensemble of N two-level atoms. It was shown that above a critical coupling strength such system can undergo QPT~\cite{wang1973phase,hepp1973superradiant,hepp1973equilibrium}, called as ``superradiant phase".  Nature of MPE entanglement, above the QPT, is quite different than the one emerging via so called spin-squeezing schemes~\cite{sorensen2001Nature,SqueezingTransferSpin_PRL_1997,Kitagawa&UedaPRA1993,Tasgin&MeystrePRA2011,MeasurementInducedSqueezing_PRA_1999,kuzmich2000generation}.   Although a bare Dicke state can be witnessed by the squeezing in the first order spin noise; such criteria fail to witness the nonclassicality of the ensemble in the superradiant phase, a superposition of many Dicke states. Hence, investigation of second-order noise becomes important~\cite{tasgin2017many}.

Recently, a Dicke-like QPT is observed also in two interacting ensembles~\cite{ZhengPRA2011ensemble_ensemble}, above a critical coupling strength. Its similarity to optical superradiance has been pointed out and the ensemble-ensemble entanglement is characterized using von Neumann entropy~\cite{ZhengPRA2011ensemble_ensemble}. Actually, such entanglement measures can be related to the global geometric entanglement~\cite{orus2008equivalence} for collective models. More detailed investigations of the quantum correlations performed subsequently in terms of scaled concurrence and quantum discord (QD)~\cite{wang2012quantum,zhang2013quantum}. It is concluded that while these criteria exhibit similar power-law scaling behaviors with that of the optical superradiance, their behavior is different from each other. In particular, QD is relatively a good measure for capturing quantum correlations compared to scaled concurrence for such superexcitation collective models ~\cite{zhang2013quantum}. However, QD counts the total amount of correlations and, hence, does not provide specific information on the existence and nature of MPE. QD is highly sensitive to system parameters, such as atomic transition frequency differences in two ensembles~\cite{zhang2013quantum}.

It is apparent that, although they are quantifications for ensemble-ensemble entanglement, these are hard to compute and measure. Thus, these criteria are not practical for experiments. Our objective here is to propose {\it experimentally-accessible} criteria to characterize MPE related with the ensemble-ensemble entanglement, and, apply those criteria to a system of two coupled atomic ensembles exhibiting a Dicke-like QPT.
  
Experimental investigations of bipartite entanglement benefit from more operational entanglement criteria, such as two-mode entanglement that can be measured by collective spin operators~\cite{giovannetti2003characterizing}. Very recently similar experiments and MPE measurements, based upon mode entanglement detections, have been reported for ultracold atomic Bose-Einstein condensates (BECs)~\cite{fadel2018spatial,kunkel2018spatially,lange2018entanglement}. There are also promising setups for producing and detecting MPE at higher temperatures. One of them is epsilon-near-zero (ENZ) materials~\cite{caglayan2017coherence} which allows for long-distance coherent interaction of two ensembles of quantum emitters. Another one is the entanglement of nano-ensembles~\cite{chen2011surface,tame2013quantum} via localized plasmons.

In this manuscript, we derive criteria for ensemble-ensemble entanglement. We use the method given in Refs.~\cite{Nha&ZubairyPRL2008,nha2007entanglement}, originally introduced for two-mode entanglement.  We realize that, in analogy with Refs.~\cite{SimonPRL2000,Agarwal&BiswasNJP2005}, where $ \hat{p}_2\rightarrow -\hat{p}_2$ in a partial transpose, collective spin operator becomes $ \hat{J}^{(y)}_2\rightarrow -\hat{J}^{(y)}_2$ in a partial transpose with respect to the second ensemble. 


 While the criterion~\cite{giovannetti2003characterizing,sorensen2001Nature}, widely used for detecting the entanglement between two atomic condensates in the experiments, adopts the uncertainty relation in first-order spin components, i.e. $ \hat{u}_1=\hat{J}_1^{(x)}+\hat{J}_2^{(x)} $~(spin squeezing-like ), our criteria look for the uncertainty relation in the second-order noise, e.g. $ \hat{u}_2=\hat{J}_1^{+} \hat{J}_2^{-}+ \hat{J}_1^{-} \hat{J}_2^{+}$~(number squeezing-like ), where $\hat{J}_i^{\pm}= \hat{J}_i^{(x)}\pm i\hat{J}_i^{(y)} $. In the thermodynamic limit, both type of criteria can successfully detect entanglement in the Dicke like quantum phase. For finite number of particles, however, we demonstrate that only the criteria based on the second-order noise can witness the entanglement in the superradiant phase.

The manuscript is organized as follows. In Sec.~\ref{sec-QPT}, we present a brief review of the results of the Ref.~\cite{ZhengPRA2011ensemble_ensemble}  by introducing the Hamiltonian for the two interacting atomic ensembles. We derive ground state wave-function of the system in the thermodynamic limit, $N\to\infty$, and show that the system can undergo a QPT at a critical coupling strength~\cite{ZhengPRA2011ensemble_ensemble}. In Sec.~\ref{sec-EC}, we obtain ensemble-ensemble entanglement criteria via using the partial transpose method. We show that entanglement accompanies the QPT in the thermodynamic limit. We also include numerical calculations for the finite number of particles, where we observe that only number squeezing-like  criteria can witness the entanglement. In Sec.~\ref{sec:BEC}, we demonstrate that the findings can be applied to a two-component condensate system trapped in a double-well potential. A  summary appears in Sec.~\ref{sec:summary}.

\section{ QUANTUM PHASE TRANSITION $\&$ ground state wavefunction} \label{sec-QPT}

In this section, we study the appearance of the Dicke-like QPT in a system of two interacting ensembles~\cite{ZhengPRA2011ensemble_ensemble} and derive the ground state wave function for the superexcitation phase by following the methods used in Ref.~\cite{emary2003chaos}. The Hamiltonian for this system can be written as~\cite{ZhengPRA2011ensemble_ensemble}
\begin{eqnarray}
\hat{\cal H} &=& \omega_{1} \hat{J}_{1}^{(z)} +\omega_{2} \hat{J}_{2}^{(z)} + \tilde{\lambda}(\hat{J}_{1}^++\hat{J}_{1}^-)(\hat{J}_{2}^++\hat{J}_{2}^-), \quad \label{Ham1}
\end{eqnarray}
where there are $ {\rm N}_i $ number of two-level atoms in the $i$th ensemble with energy level spacing $\omega_i$. Here, $ \tilde{\lambda}={\lambda}/{\sqrt{{\rm N}_1 {\rm N}_2}} $ determines the interaction strength between ensembles. $ \hat{J}_{i}^{(z)}$ and $\hat{J}_{i}^{\pm} $ are the collective spin operators for the two level atoms in the $i$th ensemble~($i=1,2$).  Collective spin operators satisfy the commutation relations $ [\hat{J}_{i}^{+},\hat{J}_{i}^{-}]=2 \hat{J}_{i}^{(z)} $ and  $ [\hat{J}_{i}^{\pm},\hat{J}_{i}^{(z)}]= \mp \hat{J}_{i}^{\pm} $. 

\begin{figure*}
\begin{center}
\includegraphics[width=170mm,height=60mm]{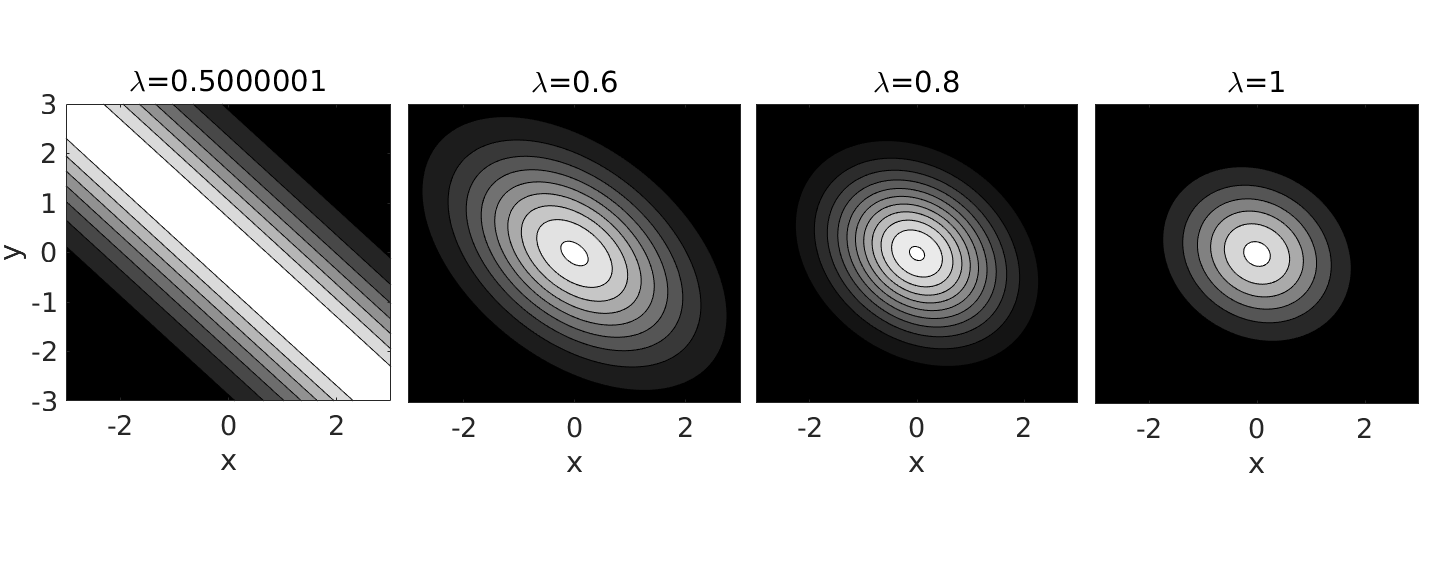}
\caption{The ground-state wave function $\Psi_{GS}^{(2)}(x,y) $ in the x-y representation for various coupling strengths $ \lambda $. The coordinates x-y can be written in terms of $ x_0 $ and $ y_0 $, representing diagonal form of Eq.~(\ref{NP_Ham}), as $ x=x_0-\sqrt{{\rm N}_1 \alpha_1} $ and  $ y=y_0-\sqrt{{\rm N}_1 \alpha_1} $. The wave function becomes stretched when the coupling strength is close to critical value, and it relaxes back to well-localized state for large coupling values~\cite{emary2003chaos}. The scaled frequencies are $ \omega_1=\omega_2=1 $ and $ \lambda_c=0.5 $.}
\label{fig-wf}
\end{center}
\end{figure*} 

By applying the Holstein-Promakoff transformation~\cite{emary2003chaos,holstein1940field,ressayre1975holstein}, the operators can be written as
\begin{eqnarray}
\hat{J}_{i}^{+}=\hat{b}_{i}^{\dagger} \sqrt{{\rm N}_i-\hat{b}_{i}^{\dagger}\hat{b}_{i}}, && \quad \hat{J}_{i}^{-}=\sqrt{{\rm N}_i-\hat{b}_{i}^{\dagger}\hat{b}_{i}} \hat{b}_{i}, \nonumber \\
\label{Hols_Prim} \\
\centering \hat{J}_{i}^{(z)}&=&\hat{b}_{i}^{\dagger}\hat{b}_{i}-{\rm N}_i/2, \nonumber
\end{eqnarray}
in terms of the annihilation operators $\hat{b}_{1,2}$. Under such a transformation, the Hamiltonian in Eq.~(\ref{Ham1}) can be written in the bosonic picture as
\begin{eqnarray}
\hat{\cal H} &=&  \omega_{1} (\hat{b}_{1}^{\dagger}\hat{b}_{1}-{\rm N}_1/2) +\omega_{2} (\hat{b}_{2}^{\dagger}\hat{b}_{2}-{\rm N}_2/2) \nonumber \\ 
\label{Bosonic_Ham} 
&+& \tilde{\lambda} (\hat{b}_{1}^{\dagger} \sqrt{{\rm N}_1-\hat{b}_{1}^{\dagger}\hat{b}_{1}}+\sqrt{{\rm N}_1-\hat{b}_{1}^{\dagger}\hat{b}_{1}} \hat{b}_{1}) \\
&\times &(\hat{b}_{2}^{\dagger} \sqrt{{\rm N}_2-\hat{b}_{2}^{\dagger}\hat{b}_{2}}+\sqrt{{\rm N}_2-\hat{b}_{2}^{\dagger}\hat{b}_{2}} \hat{b}_{2}).\nonumber
\end{eqnarray}
In the thermodynamic limit, in which the number of atoms in each ensemble becomes infinite, ${\rm N}_i\rightarrow \infty $, one can obtain the effective Hamiltonian as~\cite{ZhengPRA2011ensemble_ensemble}
\begin{eqnarray}
\hat{\cal H}^{(1)}=\sum_{i=1}^2\omega_{i} (\hat{b}_{i}^{\dagger}\hat{b}_{i}-{\rm N}_i/2) + {\lambda} (\hat{b}_{1}^{\dagger}+\hat{b}_{1})(\hat{b}_{2}^{\dagger} + \hat{b}_{2}), \label{NP_Ham}
\end{eqnarray}
which is equivalent to the spin-boson Dicke Hamiltonian in the normal phase in thermodynamic limit. If one follows the calculations of Ref.~\cite{emary2003chaos}, the excitation energies can be found as,
  \begin{eqnarray}
  \varepsilon_\pm^{2} =\Bigl[ {\omega}_1^2+{\omega}_2^2 \pm \sqrt{({\omega}_1^2-{\omega}_2^2)^2+16{\lambda}^2 \omega_1 \omega_2} \Bigr] /2. \label{NP_energy}
 \end{eqnarray}
Thus, critical coupling strength can be given as ${\lambda}_c=\sqrt{\omega_1 \omega_2}/2 $. As long as the interaction strength is smaller than this value~($ {\lambda}<{\lambda}_c $), the solutions of Eq.~(\ref{NP_energy}) are real and the system stays in the normal phase.

To find a solution above the critical point~($ {\lambda}>{\lambda}_c $), one can displace the bosonic modes with opposite signs as
 \begin{eqnarray}
\hat{b}_1^{\dagger}= \hat{d}_1^{\dagger} \pm\sqrt{{\rm N}_1 \alpha_1}, \qquad  \hat{b}_2^{\dagger}= \hat{d}_2^{\dagger} \mp\sqrt{{\rm N}_2 \alpha_2}. \label{Displacement}
\end{eqnarray}
In the following, we shall just consider the displacements as; $\hat{b}_1^{\dagger}= \hat{d}_1^{\dagger} +\sqrt{{\rm N}_1 \alpha_1}  $ and $ \hat{b}_2^{\dagger}= \hat{d}_2^{\dagger} -\sqrt{{\rm N}_2 \alpha_2} $. If we insert these definitions into Eq.~(\ref{Bosonic_Ham}) and eliminate linear terms, one can find the amounts of displacement of each mode by solving 
 \begin{eqnarray}
 \omega_1 -2\lambda \sqrt{\frac{1-\alpha_2}{1-\alpha_1}} \sqrt{\frac{{\rm N}_2\alpha_2 }{{\rm N}_1 \alpha_1}} (1-2\alpha_1)=0,  \label{a} \\ 
  \omega_2-2\lambda \sqrt{\frac{1-\alpha_1}{1-\alpha_2}} \sqrt{\frac{{\rm N}_1 \alpha_1 }{{\rm N}_2 \alpha_2}} (1-2\alpha_2)=0, 
 \label{b}
\end{eqnarray}
and obtain~\cite{ZhengPRA2011ensemble_ensemble}
\begin{eqnarray}
  \alpha_1&=& \frac{1}{2} \Bigl(1-\sqrt{\frac{4{\rm N}_1\lambda^2  \omega_1^2+{\rm N}_2 \omega_1^2\omega_2^2}{4{\rm N}_1\lambda^2  \omega_1^2+16{\rm N}_2 \lambda^4}}\Bigr), \\ 
  \alpha_2&=& \frac{1}{2} \Bigl(1-\sqrt{\frac{4 {\rm N}_2 \lambda^2  \omega_2^2 + {\rm N}_1 \omega_1^2\omega_2^2}{4{\rm N}_2 \lambda^2  \omega_2^2+16 {\rm N}_1 \lambda^4}}\Bigr). 
\end{eqnarray}
Then, the corresponding effective Hamiltonian can be found as
 \begin{eqnarray}
 \hat{\cal H}^{(2)}&=&\Omega_1 \hat{d}_1^{\dagger}  \hat{d}_1+\Omega_2 \hat{d}_2^{\dagger} \hat{d}_2+ \kappa_1 ( \hat{d}_1^{\dagger} + \hat{d}_1)^2 \nonumber \\
 &+& \kappa_2 ( \hat{d}_2^{\dagger} + \hat{d}_2)^2+\Lambda (\hat{d}_1^{\dagger} + \hat{d}_1)( \hat{d}_2^{\dagger} + \hat{d}_2),
 \label{Displaced_Ham}
\end{eqnarray}
where 
 \begin{eqnarray}
 \Omega_{1(2)}&=& \omega_{1(2)}+2\lambda \sqrt{\alpha_1 \alpha_2} \sqrt{\frac{{\rm N}_{2(1)}(1-\alpha_{2(1)})}{{\rm N}_{1(2)}(1-\alpha_{1(2)})}}, \\
 \kappa_{1(2)}&=& \lambda  \sqrt{\alpha_1 \alpha_2} \sqrt{\frac{{\rm N}_{2(1)}(1-\alpha_{2(1)})}{{\rm N}_{1(2)}(1-\alpha_{1(2)})}} \frac{2-\alpha_{1(2)}}{1-\alpha_{1(2)}}, \\ 
 \Lambda&=&\lambda \frac{(1-2 \alpha_1)(1-2 \alpha_2)}{\sqrt{(1-\alpha_1)(1-\alpha_2)}}. 
 \end{eqnarray}
The Hamiltonian in Eq.~(\ref{Displaced_Ham}) contains terms for single-mode squeezing in addition to the two-mode squeezing seen in Eq.~(\ref{NP_Ham}). These terms vanish before the critical point of QPT and do not contribute to two-mode entanglement. The diagonalization of Eq.~(\ref{Displaced_Ham}) is possible by introducing the position-momentum operators,
 \begin{eqnarray}
\hat{X}&=&  \frac{1}{\sqrt{2 \Omega_1}}  ({d}_{1}^{\dagger}+{d}_{1}), \qquad \hat{P}_X= i  \frac{\sqrt{ \Omega_1}}{2}  ({d}_{1}^{\dagger}-{d}_{1}),  \label{Eq-posx}\\
\hat{Y}&=&  \frac{1}{\sqrt{2 \Omega_2}}  ({d}_{2}^{\dagger}+{d}_{2}), \qquad \hat{P}_Y= i  \frac{\sqrt{ \Omega_2}}{2}  ({d}_{2}^{\dagger}-{d}_{2}),  \label{Eq-posy}
\end{eqnarray}  
and the eigenfrequencies can be found as,
  \begin{eqnarray}
  \epsilon_\pm^2 =\Bigl[ \tilde{\omega}_1^2+\tilde{\omega}_2^2 \pm \sqrt{(\tilde{\omega}_1^2-\tilde{\omega}_2^2)^2+16{\Lambda}^2 \Omega_1 \Omega_2} \Bigr] /2,
 \end{eqnarray}
where $ \tilde{\omega}_i^2= {\Omega}_i( \Omega_i + 4 {\kappa}_i)$. Thus, in this representation in which the effective Hamiltonian is diagonal, its ground state wave function can simply be given as~\cite{emary2003chaos},
\begin{widetext}
\begin{equation}
 \Psi_{GS}^{(2)}(x,y)=\Bigl(\frac{\epsilon_- \epsilon_+}{\pi^2}\Bigr)^{1/4} exp\Bigl\{ -\frac{\epsilon_-}{2}[x \cos(\theta)-y \sin(\theta)]^2 -\frac{\epsilon_+}{2}[x \sin(\theta)+y \cos(\theta)]^2  \Bigl\},\label{Eq-GsWF}
\end{equation}
\end{widetext}
here $\theta={\rm tan^{-1}}[{4\Lambda \sqrt{\Omega_1 \Omega_2}} /({\tilde{\omega}_1^2-\tilde{\omega}_2^2})]/2$, and we define $ x_i=\sqrt{\omega_i/\Omega_i}X_i$ for the $i$th ensemble (i=1,2) with $x_{1,2}=x,y$. The plot of x-y representations of the ground state wavefunction is given in FIG.~\ref{fig-wf} for different coupling strengths, $ \lambda $. In this phase, the wave function becomes stretched when the coupling strength is close to the critical value. As the coupling increases, $ \lambda\gg \lambda_c $, it relaxes back to a localized state, which is a similar behavior to the one obtained in Ref.~\cite{emary2003chaos} for the Dicke model.

\begin{figure}
\begin{center}
\includegraphics[width=90mm,height=35mm]{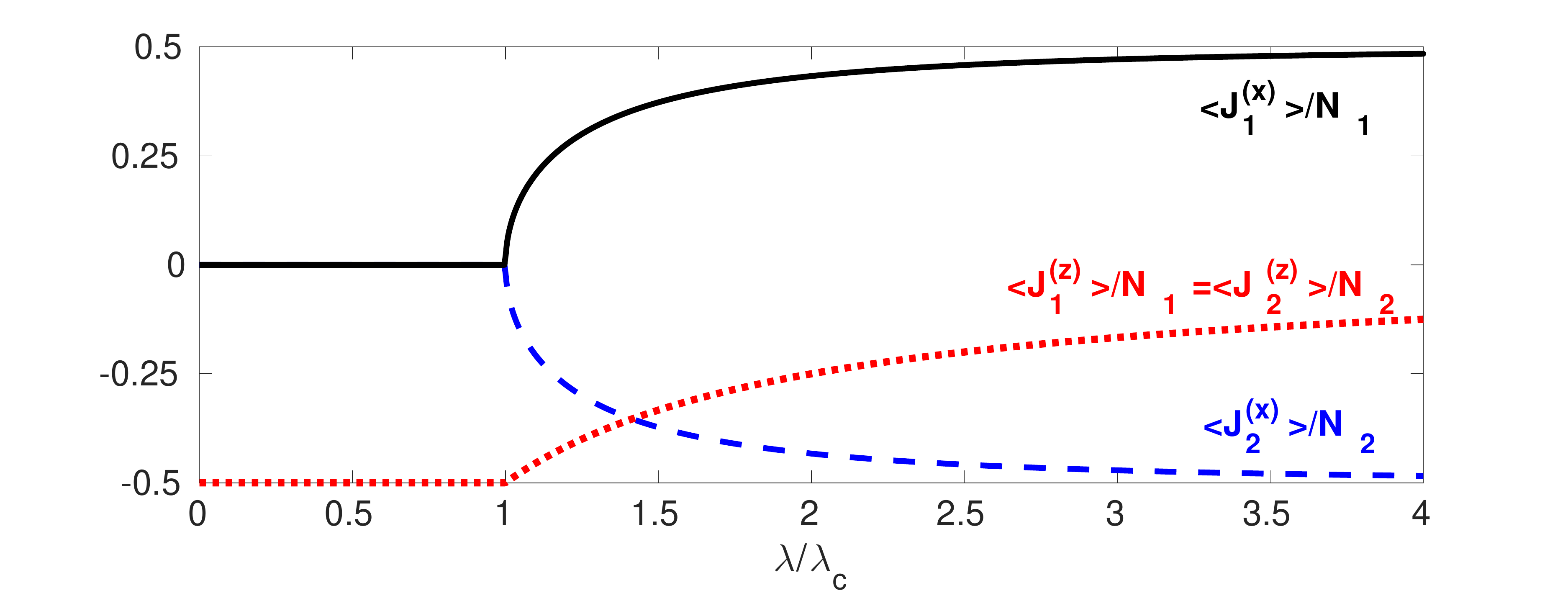}
\caption{Expectation values of angular momentum operators in the thermodynamic limit as a function of coupling strength. The resonance frequencies are taken as $ \omega_1=\omega_2=1 $ and we use $ {\rm N}_1 $ = $ {\rm N}_2 $ = N .}
\label{fig_jexp}
\end{center}
\end{figure} 
In the thermodynamic limit, the collective angular momentum operators can be given by [see Appendix]
\begin{eqnarray}
\hat{J}_i^{(x)}&\cong & \chi_i {\rm N}_i \sqrt{ \alpha_i (1-\alpha_i)}+\sqrt{\frac{{\rm N}_i \Omega_i}{2}} \frac{1-2\alpha_i}{\sqrt{1-\alpha_i}}\hat{X}_i, \label{Eq:Jx}\\
\hat{J}_i^{(y)}& \cong &-\sqrt{\frac{{\rm N}_i (1-\alpha_i) }{2 \Omega_i}}\hat{P}_{X_i}, \label{Eq:Jy} \\
\hat{J}_{i}^{(z)}&\cong &{\rm N}_i (\alpha_i-1/2)+\chi_i \sqrt{2{\rm N}_i \Omega_i \alpha_i} \hat{X}_i,  \label{Eq:Jz}
\end{eqnarray}
where $   \chi_{1}= 1$,  $\chi_{2}= -1$ and $X_{1,2}=X,Y$. In the ground state, $ \langle \hat{J}_i^{(y)}\rangle =0$ and the expectation values of the other components can be obtained as
\begin{equation}
\frac{\langle \hat{J}_i^{(z)}\rangle}{N_i}=\begin{cases}
-0.5, &  \lambda\leq \lambda_c\\
(\alpha_i-0.5), & \lambda> \lambda_c
\end{cases}
\label{Eq-Jz}
\end{equation}
and
\begin{equation}
\frac{\langle \hat{J}_i^{(x)}\rangle }{N_i}=\begin{cases}
0, &  \lambda\leq \lambda_c\\
\chi_i \sqrt{ \alpha_i (1-\alpha_i)}, & \lambda> \lambda_c
\end{cases}
\end{equation}
in which one can clearly observe that above $ \lambda_c $ each atomic ensemble acquire macroscopic excitations with finite  and macroscopically large atomic polarization~($\hat{J}_i^{(x)}=N_i \chi_i \sqrt{ \alpha_i (1-\alpha_i)} $). This breaks parity symmetry of the system accordingly, and it is also related with that the ground state becomes two-fold degenerate~\cite{ZhengPRA2011ensemble_ensemble}. In FIG.~\ref{fig_jexp}, we plot the expectation values of the collective angular momentum operators as a function of coupling strength. It can be seen from the figure that the atomic inversion~($ {\langle \hat{J}_i^{(z)}\rangle }/{\rm N}_i $) in each ensemble increases when coupling strength exceeds critical coupling. The values of polarization in each ensemble~($ {\langle \hat{J}_i^{(x)}\rangle }/{\rm N}_i $) increases with increasing interaction with opposite sign.

\section{Ensemble-Ensemble Entanglement  Criteria} \label{sec-EC}
It is known that atomic~(or spin) coherent states mimic the coherent states of a single-mode light~\cite{klauder1985applications,radcliffe1971JPhysA} as the number of particles in the ensemble becomes very large. Ref.~\cite{tasgin2017many} uses this relation to demonstrate that a MPE criterion can be converted into a single-mode nonclassicality~(SMNc) condition. (i) In Ref.~\cite{tasgin2015HP}, it is shown that quadrature-squeezing (a SMNc) condition can be obtained from the spin-squeezing criterion~\cite{sorensen2001Nature}. That is, MPE in a spin-squeezed system is associated with a quadrature-squeezing in the quasiparticle excitations of the system. (ii) Ref.~\cite{tasgin2017many} utilizes such a connection ---between the atomic coherent states and the coherent states of light--- to construct a trial form for a many-particle entanglement criterion and show that such a criterion works well in the superradiant phase. The following question is raised. If the spin-squeezing criterion, measures the noise in the first-order spin operators, which converges to quadrature-squeezing condition, e.g., $ \hat{J}_x\rightarrow x $; what kind of a MPE criterion should converge to number-squeezing (a SMNc) condition (also known as Mendel' s Q-parameter)? The simplest choice for such an operator would be $\Delta (\hat{J}_+ \hat{J}_-) \rightarrow \Delta \hat{n}$, i.e., as the large $N$ limit of the Holstein-Primakoff transformation~\cite{emary2003chaos,holstein1940field}.  Examining the uncertainty of $\hat{\cal R}=\hat{J}_+ \hat{J}_-$ Ref.~\cite{tasgin2017many} manages to obtain a new MPE criterion. This criterion witnesses the MPE of a single ensemble in the superradiant phase. Hence, MPE in the superradiant phase can be associated to the reduction in the number-fluctuations of the quasiparticle excitations of the ensemble~\footnote{ A reduction in the second-order noise of the quasiparticles of the ensemble is, actually, predicted from the number-squeezed form of the superradiantly emitted light~\cite{ye2011superradiance,tasgin2015measure}}.

In this work, we consider a similar mapping between two ensembles and the two-mode system. In the limit of large particle numbers, $N_{1,2}\to\infty$, the spin operators become, e.g., $\hat{J}_i^+\to\sqrt{N_i}\hat{a}_i^\dagger$. Hence, in this limit, entanglement between the two ensembles can also be witnessed by the entanglement of the $\hat{a}_{1,2}$ (mode) operators, which actually, represent the quasiparticle excitations of the ensembles, respectively~\footnote{Actually, it would not be too hard to see that an ensemble-ensemble entanglement criterion becomes a two-mode entanglement criterion in the infinite particle numbers limit using the arguments Ref.~\cite{tasgin2017many} works.}.  
We first construct a trial form, e.g. $\hat{A}_{1,2}$ and $\hat{B}_{1,2}$ below, for the collective spin operators of the two ensembles. Then, 
 similar to references~\cite{Nha&ZubairyPRL2008,Agarwal&BiswasNJP2005,SimonPRL2000}, employing the partial transpose method ($\hat{p}_2\to-\hat{p}_2$) in their approach, we also use $\hat{J}_y^{(2)} \to -\hat{J}_y^{(2)}$ in our treatment. We, in advance, note that the validity of the derivations we carry out are not related with the method we construct our trial operators whose uncertainties are examined. We merely choose the form of the noise to be investigated via such a method.

\begin{figure}
\begin{center}
\includegraphics[width=100mm,height=35mm]{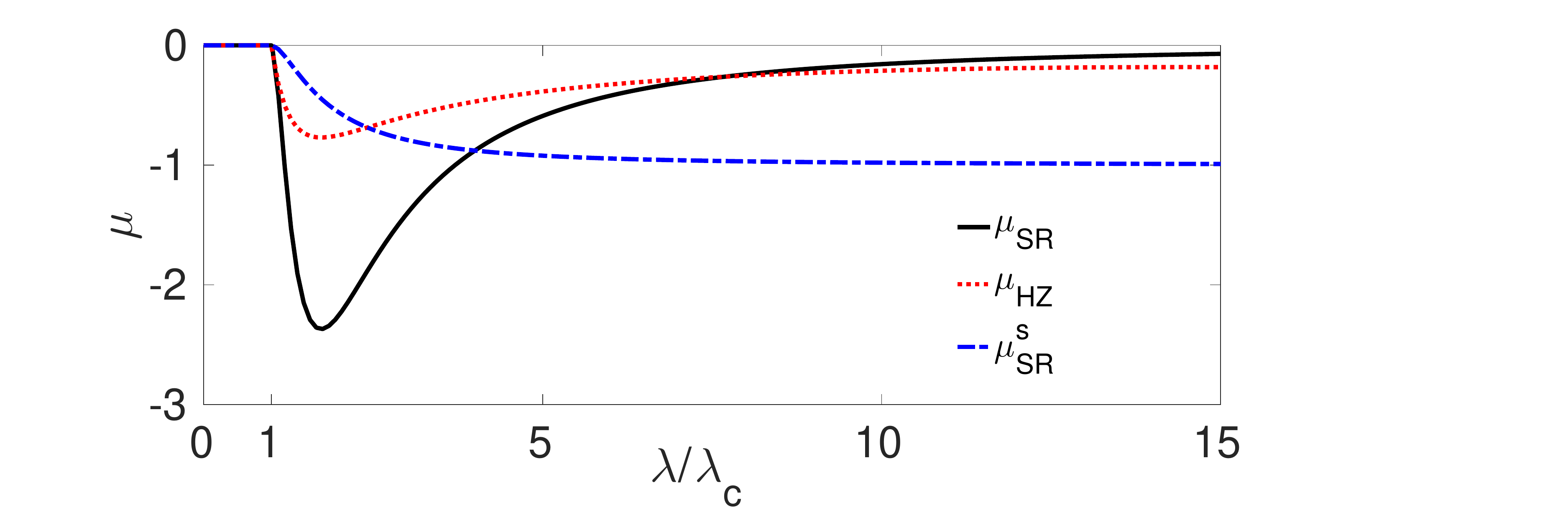}
\caption{Ensemble-ensemble entanglement in the ground state of two coupled atomic ensemble for the infinite number of particles as a function of coupling strength $\lambda$. The results are demonstrated for $\mu_{_{\rm SR}}$~(black-solid line), $\mu_{_{\rm HZ}}$~(red-dotted line), and  $\mu_{_{\rm SR}}^{s}$~(blu-dashed line). The entanglement is present for the values $\mu<0$. Here, we scale angular momentum operators with J=N/2, i.e $ \hat{J}^\pm_i \rightarrow \hat{J}^\pm_i /{\rm J}$ and we use scaled resonance frequencies as $ \omega_1=\omega_2 $ = 1.}
\label{fig-criteria}
\end{center}
\end{figure}

In the following, we derive three criteria for the ensemble-ensemble entanglement. The form of the first one is constructed by examining a quadrature-squeezing two-mode entanglement criterion, Duan-Giedke-Cirac-Zoller~(DGCZ) criterion~\cite{DGCZ_PRL2000}, which adapts  $\hat{u}=\hat{x}_1+\hat{x}_2$ and $\hat{v}=\hat{p}_1-\hat{p}_2$, and derived by applying the partial transpose ($\hat{J}_y^{(2)} \to -\hat{J}_y^{(2)}$) method. The form of the second one is constructed by examining the Schr\"{o}dinger-Robertson~(SR), stronger form of the Hillery-Zubairy~(HZ) criterion~\cite{Nha&ZubairyPRL2008}, and derived by applying the partial transpose method. A third one is obtained from the HZ criterion~\cite{Hillery&ZubairyPRL2006} by merely applying the Cauchy-Schwartz inequality. The second and the third criterion investigates the second-order noises of spin operators, while the first one deals only with the first-order noises, see the Appendix. We note that criteria obtained using the SR inequality are stronger (definite) because SR inequality also takes the intra-mode rotations into account partially, see Sec.~II.4 and Sec.~III.2 in Ref.~\cite{tasgin2019anatomy}.

We note that criteria obtained from the squeezing of the spin noise and the ones via squeezing of the bilinear products of the collective spin operators are successful in different regimes of the entanglement. The former ones are more useful when the standard quadrature or spin squeezing Hamiltonians are involved in the generation of the ensemble-ensemble entanglement. While the latter ones are more useful for strong coupling regime where an ensemble-ensemble entanglement is associated with a phase transition like structure~\cite{tasgin2017many,nha2006entanglement,vogel2014unified}. 

We first introduce the SR inequality. For a given two non-commuting observables~$ \{\hat{\rm A},\hat{\rm B} \}$, SR inequality can be written as~\cite{SR,robertson1934indeterminacy}
 \begin{eqnarray}
\langle(\Delta \hat{\rm A})^2\rangle \langle(\Delta \hat{\rm B})^2\rangle \geq \frac{1}{4} |\langle[\hat{\rm A},\hat{\rm B}]\rangle|^2 + \langle \Delta \hat{\rm A} \Delta \hat{\rm B} \rangle_S^2, \label{Eq-Sr-ine}
 \end{eqnarray}
where $ \langle \Delta \hat{\rm A} \Delta \hat{\rm B} \rangle_S=\frac{1}{2} \{\Delta \hat{\rm A} \Delta \hat{\rm B}+ \Delta \hat{\rm B}\Delta \hat{\rm A} \}$ 
defines symmetric part and when it vanishes, the inequality reduces to Heisenberg Uncertainty Relation~(HUR). Therefore, SR inequality provides a stronger uncertainty relation than HUR. In the thermodynamic limit, the symmetric part does not affect the results of this paper. However, it can play significant role in a system combined of different parameters. If we define observables $ \hat{\rm A}_1=\hat{J}^+_1 \hat{J}^+_2 +\hat{J}^-_1 \hat{J}^-_2$,
 $ \hat{\rm B}_1=i(\hat{J}^+_1 \hat{J}^+_2-\hat{J}^-_1 \hat{J}^-_2 )$  for two-mode number squeezing-like, 
 $ \hat{\rm A}_2=\hat{J}_1^{(x)}+\hat{J}_2^{(x)} $  and 
 $ \hat{\rm B}_2=\hat{J}_1^{(y)}+\hat{J}_2^{(y)} $ for two mode spin squeezing-like, under partial transpose, SR inequality  can be obtained as; $ \langle(\Delta \hat{\rm A}_i)^2\rangle_{_{\rm PT}} \langle(\Delta \hat{\rm B}_i)^2\rangle_{_{\rm PT}} \geq \frac{1}{4} |\langle[\hat{\rm A}_i,\hat{\rm B}_i]\rangle_{_{\rm PT}}|^2 +  \langle \Delta \hat{\rm A}_i \Delta \hat{\rm B}_i \rangle_{_{\rm S, PT}}^2 $~\cite{NhaPRA2006Fock_states,Nha&ZubairyPRL2008}. Therefore, the entanglement criteria can be found as~[see Appendix] 
 \begin{eqnarray}
\mu_{_{\rm SR}}&=&\langle(\Delta \hat{\rm A}_1)^2\rangle_{_{\rm PT}} \langle(\Delta \hat{\rm B}_1)^2\rangle_{_{\rm PT}}\nonumber \\
&-&\frac{1}{4} |\langle[\hat{\rm A}_1,\hat{\rm B}_1]\rangle_{_{\rm PT}}|^2 -\langle \Delta \hat{\rm A}_1 \Delta \hat{\rm B}_1 \rangle_{_{\rm S, PT}}^2 \geq 0,\label{Eq-SR}\\
\mu_{_{\rm HZ}}&=& \langle \hat{J}^+_1 \hat{J}^-_1 \hat{J}^+_2 \hat{J}^-_2\rangle-|\langle \hat{J}^+_1\hat{J}^-_2\rangle|^2 \geq 0, \label{Eq-HZ}\\
\mu_{_{\rm SR}}^{s}&=&\langle(\Delta \hat{\rm A}_2)^2\rangle_{_{\rm PT}} \langle(\Delta \hat{\rm B}_2)^2\rangle_{_{\rm PT}}\nonumber \\
&-&\frac{1}{4} |\langle[\hat{\rm A}_2,\hat{\rm B}_2]\rangle_{_{\rm PT}}|^2 -\langle \Delta \hat{\rm A}_2 \Delta \hat{\rm B}_2 \rangle_{_{\rm S, PT}}^2 \geq 0. \label{Eq-SR-ss}
 \end{eqnarray}
The PT in Eq.~(\ref{Eq-SR}) and Eq.~(\ref{Eq-SR-ss}) should yield physical density matrices if the two modes are separable. Hence, SR inequality needs to be satisfied also for the partial transposed system. Instead of evaluating the PT density matrix, one can alternatively~\cite{Agarwal&BiswasNJP2005,Nha&ZubairyPRL2008,nha2007entanglement} put $\hat{J}_y\to -\hat{J}_y$.~\footnote{Collective spin operators display a similar behavior to the quadrature operators $\hat{x}$ and $\hat{y}$. This similarity becomes exact when the ensemble has an infinite number of particles~\cite{tasgin2017many}. Using this relation, one can even deduce single-mode nonclassicality criteria from a MPE entanglement criteria.}

 The uncertainties of the operators $\hat{A}_{1,2}$ and $\hat{B}_{1,2}$ can be represented in terms of the components of the collective spin operators~[see Eq.(\ref{components})]. These quantities are measurable in the experiments~\cite{lange2018entanglement,gross2010nonlinear,measure_higher-order-polarization-moments,suter1994experimental}.

The violation of the inequality,~$ \mu <$ 0, experiences the presence of an ensemble-ensemble entanglement between the two collective spins. From the differentiation in the form of the two operator sets, $\hat{A}_1$, $\hat{B}_1$ and $\hat{A}_2$, $\hat{B}_2$, one can realize that they work better for different kind of states. While $\mu^s_{SR}$ refers to the criterion for spin squeezed-like states,  $\mu_{SR}$ refers to the criterion for the number squeezed-like states. 

We demonstrate the results of the Eqs.~(\ref{Eq-SR})-(\ref{Eq-SR-ss}) in FIG.~\ref{fig-criteria} with respect to the coupling strength. When the coupling strength between the two ensembles exceeds the critical value for the QPT, there appears a transition also in the entanglement. The strength of the violation of $\mu$ accompanies the QPT. In the symmetry-breaking phase ($ \lambda>\lambda_c $), number squeezing-like  criteria ($ \mu_{SR} $ and $ \mu_{HZ} $) take larger values in the mediate coupling regime above the critical point. As the coupling strength increases, they decrease monotonically and approach zero in the strong coupling limit. Whereas, spin squeezing-like  criterion ($ \mu^s_{SR} $) works better in the strong-coupling limit~[see FIG.~\ref{fig-criteria}]. It is important to note that these values do not reflect the strength of the entanglement, but give an idea about behaviors of the criteria.

\begin{figure}
\begin{center}
\includegraphics[width=100mm,height=35mm]{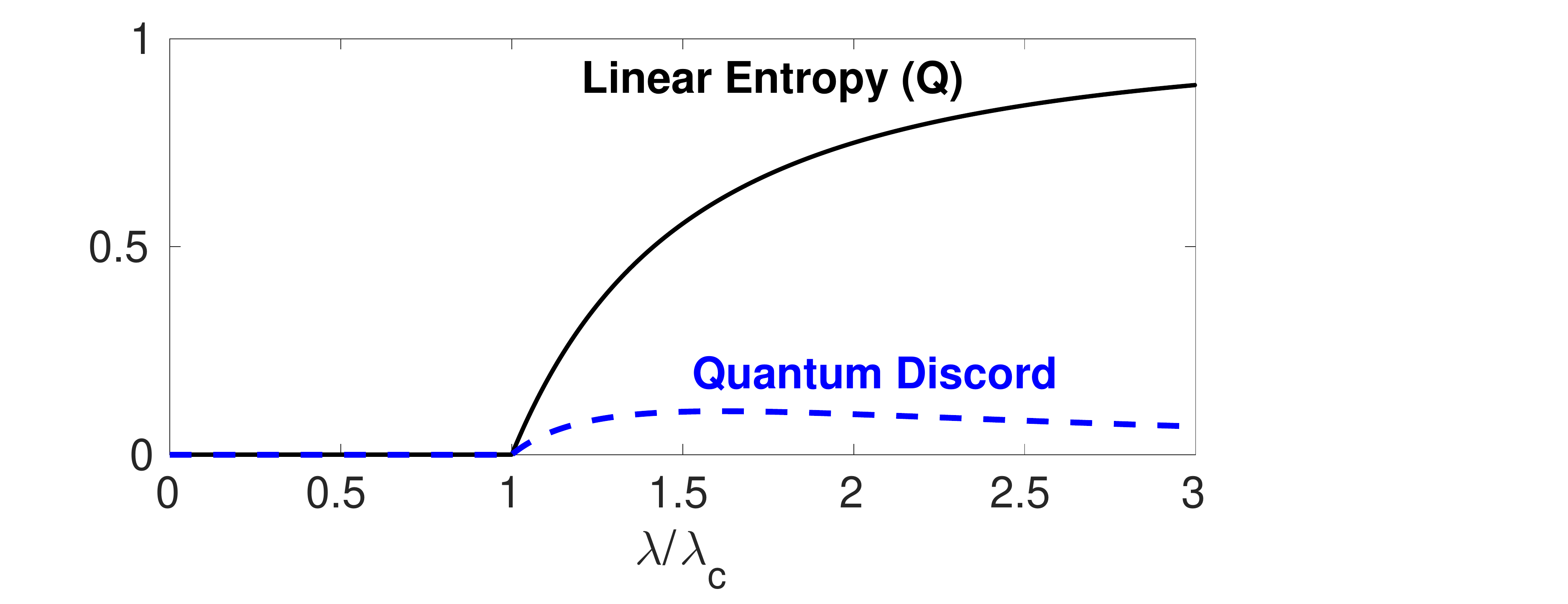}
\caption{The linear entropy (black-straight line) and QD (blue-dashed line) in different phases as a function of coupling strength $ \lambda $, in the thermodynamic limit.}
\label{fig-entropy}
\end{center}
\end{figure}

To make our results more relevant, in FIG.~\ref{fig-entropy} we reproduce linear entropy~(Q)~\cite{lambert2005entanglement} and QD~\cite{wang2012quantum,zhang2013quantum} calculations. Like our criteria, these quantities also display a clear discontinuity at the critical point. Interestingly, the behaviors of the linear entropy and the spin squeezing-like criterion, $ \mu^s_{SR} $, are similar. They both increase with increasing interaction. Whereas number squeezing-like  criteria, $ \mu_{SR} $ and $ \mu_{HZ} $, and QD decay at larger interactions.
 
\begin{figure}
\begin{center}
\includegraphics[width=100mm,height=35mm]{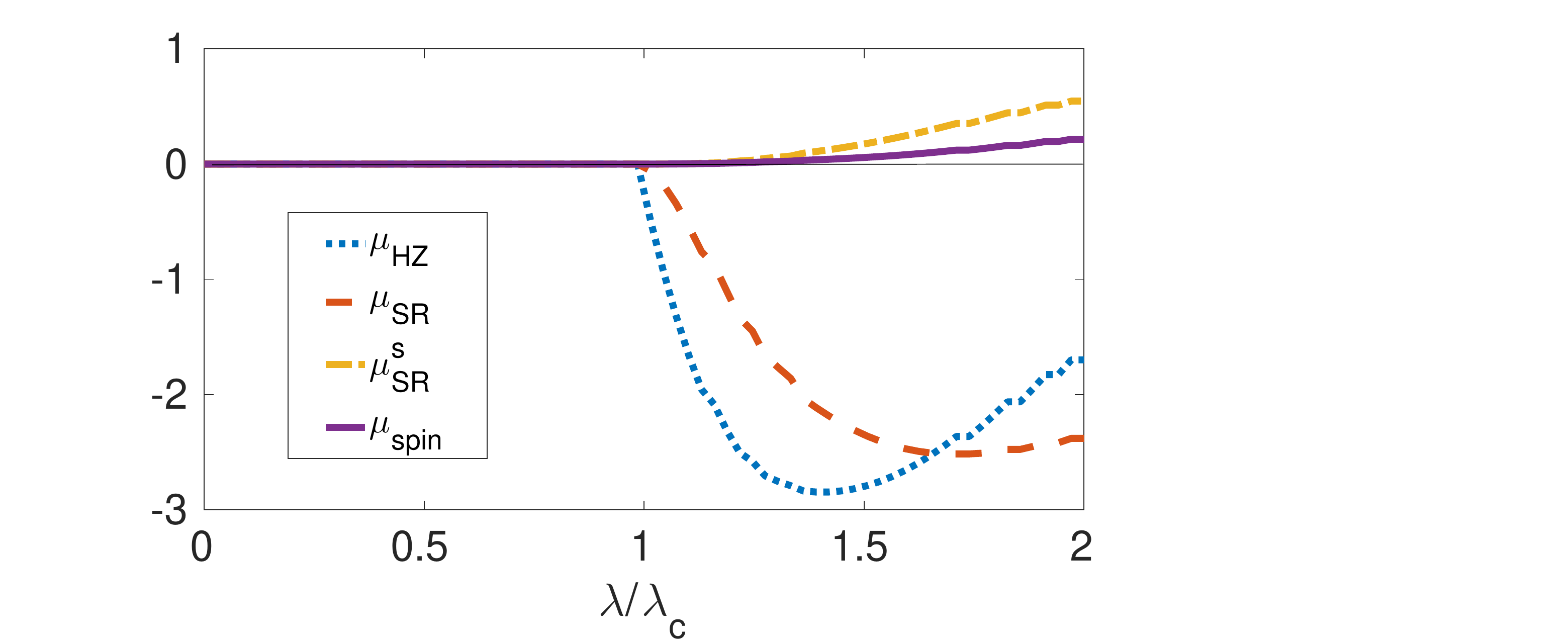}
\caption{Ensemble-ensemble entanglement in the ground state of two coupled atomic ensemble for finite number of particles, $ {\rm N}_1={\rm N}_2=40 $, as a function of coupling strength $\lambda$. The results for number squeezing-like  criteria, $\mu_{_{\rm SR}}$, $\mu_{_{\rm HZ}} < 0 $ witness entanglement, in which the spin squeezing-like  criteria, $\mu_{_{\rm SR}}^{s} $ and $\mu_{\rm spin} $ fails to detect.}  
\label{fig-finite}
\end{center}
\end{figure}

In the case of the finite number of particles, where Holstein-Primakoff transformation fails and one has to carry the calculations numerically, spin squeezing based criterion~\cite{sorensen2001Nature,giovannetti2003characterizing} cannot witness the entanglement. In FIG.~\ref{fig-finite}, we compare the results of number squeezing-like , $\mu_{_{\rm SR}}$, $\mu_{_{\rm HZ}}$ and spin squeezing-like ,  $\mu_{_{\rm SR}}^{s} $, $\mu_{\rm spin} $ criteria for finite number of particles, where we take $ {\rm N}_1={\rm N}_2=40 $. We define the criterion, $\mu_{\rm spin} $, as~\cite{giovannetti2003characterizing}
 \begin{eqnarray}
\mu_{_{\rm spin}}&=&\langle(\Delta \hat{\rm A}_2)^2\rangle \langle(\Delta \hat{\rm B}_2)^2\rangle-\frac{1}{4} |\langle[\hat{\rm A}_2,\hat{\rm B}_2]\rangle |^2 . \label{Eq-spin}
 \end{eqnarray}
It is apparent from FIG.~\ref{fig-finite} that squeezing in the fluctuations of the number of quasiparticle excitations is different than squeezing in the spin fluctuations and MPE may not be captured from spin squeezing-like ,  $\mu_{_{\rm SR}}^{s} $, $\mu_{\rm spin} $ criteria for small number of particles. 

\begin{figure*}
\begin{center}
\includegraphics[width=150mm,height=50mm]{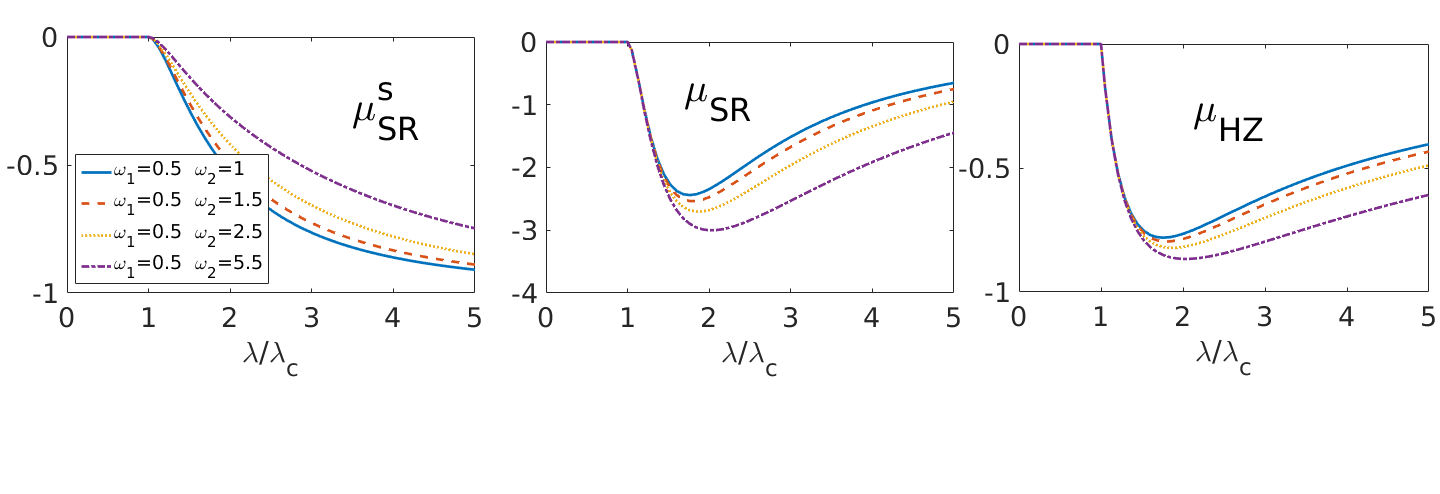}
\caption{The results for $\mu_{_{\rm SR}}^{s} $, $\mu_{_{\rm SR}}$ and $\mu_{_{\rm HZ}}$  for different values of $\omega_2$ = 1, 1.5, 2.5, 5.5 and fixed $\omega_1$ = 0.5 in the thermodynamic limit. $ \mu < 0 $ indicates the presence of an ensemble-ensemble entanglement.}
 \vskip-0.09truecm
\label{fig-mu_BEC}
\end{center}
\end{figure*}

This differentiation can play significant role, especially, in the area of quantum plasmonics, where two or more nano-diamonds including tens of color centers can be entangled via strong effective coupling provided by the common plasmon field. Metal nanoparticles localize incident field on the nanometer dimensions, which creates an extreme plasmon-color center interaction. This interaction leads to correlations between the two ensembles (of color centers) in the effective strong coupling regime~\cite{yang2015dynamics}. Interestingly, plasmons can support nonclassical features, related with noise operators~\cite{tame2013quantum}, much longer times compared to the amplitude damping rates~\cite{varro2011hanbury}.

\section{Bose-Einstein condensate in a double-well potential} \label{sec:BEC}
The physical realization of entanglement between two atomic ensembles can be found in a two-component condensates trapped in a double-well potential. It was shown that entanglement, in Bose-Einstein condensate~(BEC) systems, can be generated with the mixture of two kinds of atoms~\cite{simoni2003magnetic,roy2017two} or on the sublevels of the same kind of atoms~\cite{myatt1997production,hall1998dynamics}. Moreover, it is possible to control interactions externally~\cite{simoni2003magnetic,thalhammer2008double}, which makes these systems one of the most appropriate candidates for testing criteria defined in the previous section. In the following, we demonstrate that the Hamiltonian in Eq.~(\ref{Displaced_Ham}) can be obtained in a two-component BEC structure trapped in a double-well potential.

Let us consider that the condensates trapped deeply in a double-well potential. To observe entanglement between these sites, we use two modes $ \hat{a}_1 $ ($  \hat{b}_1 $) and $ \hat{a}_2 $ ($  \hat{b}_2 $) for the first (second) site. These modes can be spin components of the same well. Total Hamiltonian of such system can be given as~\cite{ng2005two} 
 \begin{eqnarray}
\hat{\cal H} &=& \frac{ \omega_1}{2} (\hat{a}_1^\dagger  \hat{b}_1+\hat{b}_1^\dagger  \hat{a}_1)+ \frac{ \omega_2}{2} (\hat{a}_2^\dagger  \hat{b}_2+\hat{b}_2^\dagger  \hat{a}_2) \nonumber\\
&+& \frac{ \tilde{\kappa}_1  }{2} [(\hat{a}_1^\dagger  \hat{a}_1 )^2 +( \hat{b}_1^\dagger  \hat{b}_1  )^2  ] +\frac{ \tilde{\kappa}_2  }{2} [(\hat{a}_2^\dagger  \hat{a}_2 )^2 +( \hat{b}_2^\dagger  \hat{b}_2  )^2  ] \nonumber \\
&+&\tilde{\lambda} (\hat{a}_1^\dagger  \hat{a}_1 \hat{b}_1^\dagger  \hat{b}_1+\hat{a}_2^\dagger  \hat{a}_2 \hat{b}_2^\dagger  \hat{b}_2). \label{Eq:Ham_BEC}
 \end{eqnarray}
Here,  $ \omega_i $,  $ \tilde{\kappa}_i= {\kappa}_i/{\rm N}_i$ and  $  \tilde{\lambda}={\lambda}/{\sqrt{{\rm N}_1 {\rm N}_2}} $  are the tunneling, intra-species and interspecies interaction strengths of the $i$th site respectively with  $i$=1,2. If we define the intrawell spin operators as~\cite{he2012einstein,ng2005two} :
 \begin{eqnarray}
\hat{J}_i^{(x)}&=& (\hat{a}_i^\dagger  \hat{a}_i-\hat{b}_i^\dagger  \hat{b}_i)/2,\\
\hat{J}_i^{(y)}&=&(\hat{a}_i^\dagger  \hat{b}_i-\hat{b}_i^\dagger  \hat{a}_i)/2i, \\
\hat{J}_{i}^{(z)}&=&(\hat{a}_i^\dagger  \hat{b}_i+\hat{b}_i^\dagger  \hat{a}_i)/2,  \qquad i:1, 2.
 \end{eqnarray}
Then, Eq.~(\ref{Eq:Ham_BEC}) can be written in terms of spin operators as~\cite{ng2005two}:
 \begin{eqnarray}
\hat{\cal H} &=& \sum_{i=1,2} \{\omega_i \hat{J}_i^{(z)}+\tilde{\kappa}_i  \hat{J}_i^{(x)^2}\}+ \tilde{\lambda} \hat{J}_1^{(x)}\hat{J}_2^{(x)}. \label{Ham:nonlinear}
 \end{eqnarray}
When the intra-species interactions, $\tilde{\kappa}_i$, are negligible, this Hamiltonian reduces to the one given in Eq. ~(\ref{Ham1}). For the symmetric case, having the same kinds of atoms in both sites, the previous results are valid. However, in general, one can expect different physical parameters for the two ensembles composed of different kinds of atoms~\cite{modugno2002two}, e.g. in one site Rb-atoms and in another site K-atoms. In  FIG.~\ref{fig-mu_BEC}, we demonstrate the behaviors of the Eqs.(\ref{Eq-SR}-\ref{Eq-SR-ss}) by using different values of $ \omega_1 $ and $\omega_2$. Here, one can observe that entanglement criteria still accompany the QPT.

In the presence of intra-species interactions, the Hamiltonian in Eq.~(\ref{Ham:nonlinear}) is still diagonalizable. By applying Holstein-Primakoff transformation as defined in Eq.~(\ref{Hols_Prim}), it can be written as:
\begin{eqnarray}
\hat{\cal H} &=& \sum_{i=1,2} \{ \omega_{i} (\hat{b}_{i}^{\dagger}\hat{b}_{i}-{\rm N}_i/2)  \nonumber \\ 
&+& \tilde{\kappa}_i(\hat{b}_{i}^{\dagger} \sqrt{{\rm N}_i-\hat{b}_{i}^{\dagger}\hat{b}_{i}}+\sqrt{{\rm N}_i-\hat{b}_{i}^{\dagger}\hat{b}_{i}} \hat{b}_{i})^2\} \nonumber \\
&+& \tilde{\lambda} (\hat{b}_{1}^{\dagger} \sqrt{{\rm N}_1-\hat{b}_{1}^{\dagger}\hat{b}_{1}}+\sqrt{{\rm N}_1-\hat{b}_{1}^{\dagger}\hat{b}_{1}} \hat{b}_{1}) \nonumber \\
&*&(\hat{b}_{2}^{\dagger} \sqrt{{\rm N}_2-\hat{b}_{2}^{\dagger}\hat{b}_{2}}+\sqrt{{\rm N}_2-\hat{b}_{2}^{\dagger}\hat{b}_{2}} \hat{b}_{2}).\label{Bosonic_Ham_BEC} 
\end{eqnarray}
In the thermodynamic limit, by following the calculations in Sec. II, the critical coupling strength can be obtained as: $ \lambda_c=\sqrt{(\omega_1+4 \kappa_1)(\omega_2+4 \kappa_2)} /2$. Above this critical value, $ \lambda>\lambda_c $, one can reach Eq.~(\ref{Displaced_Ham}) with modified parameters~\cite{gunay2019binary}. 

\section{Summary} \label{sec:summary}
In summary, we obtain ensemble-ensemble entanglement criteria with the use of a partial transpose method, which was originally introduced to detect two-mode entanglement~\cite{Nha&ZubairyPRL2008,Agarwal&BiswasNJP2005}. The derivations rely on that: two ensembles in a separable state satisfy the Schrodinger-Robertson inequalities even one employs the partial transpose to the state of one of the ensembles. Reduction of a noise product below a critical value indicates the ensemble-ensemble entanglement. Hence, entanglement witnesses can be associated with a specific type of noise. We show that, while a reduced noise in the first-order collective spin operators can indicate the entanglement of, e.g., between the two components of a condensate for a fair coupling~\cite{simoni2003magnetic,roy2017two}, the examination of the second-order moments of the collective spins can be required for different setups. For instance, in the case of a small number of particles, we find that the criteria using first (second)-order noise fails (succeeds) to detect the entanglement. This result makes our findings significant for the experimental detection of entanglement in a system of a small number of particles. The criteria, given in this manuscript, can be practiced in the new-generation setups operating at room temperature where, for instance, color centers of two nanodiamonds can be entangled via the strong coupling provided by the common plasmon field~\cite{chen2011surface,tame2013quantum}. Localized plasmons can couple to the quantum emitters with orders of magnitude stronger compared to the coupling of the same emitters to the plane-waves. Moreover, we also discuss that the physical realizations of the model, studied here, can be found in a two-component condensate trapped in a double-well potential.

Besides their implementations, the structures of the ensemble-ensemble entanglement criteria and their relations with the two-mode entanglement can shed light on the entanglement characterization for interacting atomic ensembles. These systems offer a notable potential for quantum sensing, quantum memory, and quantum heat engine applications.

\section*{Acknowledgments}
We thank Julien Vidal for fruitful comments. MG and MET acknowledge support from TUBITAK Grant No. 117F118. OEM and MET acknowledges support from TUBITAK-COST Grant No. 116F303. MET acknowledges support from TUBA-GEBIP 2017 fund.
\appendix
\renewcommand{\thesubsection}{\Alph{subsection}}

\section*{Appendices}

We start by deriving a criterion based on the collective spin noise using the method, $J_y^{(2)} \to - J_y^{(2)}$ under a partial transposition. We use the form of the Duan-Giedke-Cirac-Zoller two-mode entanglement criterion~\cite{DGCZ_PRL2000}. This criterion, then, has a nature similar to the one~\cite{giovannetti2003characterizing,sorensen2001Nature} widely used for multi-component condensates as given in Eq.~(\ref{Eq-spin}). Next, we derive an ensemble-ensemble entanglement criterion which is based on the measurements of the bilinear products of the collective spin operators. These criteria are obtained, again using the new method, but this time performing $J_y^{(2)} \to - J_y^{(2)}$ transform in the SR version of the higher-order spin operators, i.e. $\hat{\rm A}_1$. The form of $\hat{\rm A}_1$ is constructed by examining the form of the SR version~\cite{Nha&ZubairyPRL2008} of the HZ criterion~\cite{Hillery&ZubairyPRL2006}. The number squeezing-like  criteria, although can measure the ensemble-ensemble entanglement in the superradiant phase, necessitate the measurement of the second-order moments of the spin components~\cite{measure_higher-order-polarization-moments}.

\subsection{Derivation of Eqs.(\ref{Eq-SR})-(\ref{Eq-SR-ss})}
One can derive inequalities defined in Eq.~(\ref{Eq-SR}) and Eq.~(\ref{Eq-SR-ss}) by using the method introduced in Ref's.\cite{NhaPRA2006Fock_states,nha2007entanglement}, which are based on the violation of SR inequalities~\cite{Nha&ZubairyPRL2008,nha2007entanglement}. To do this,  we can transform operators~\cite{tasgin2017many} as  $ \hat{a}_i\rightarrow \hat{J}^-_i $ and $ \hat{a}_i^\dagger \rightarrow \hat{J}^+_i $. If we define observables as
\begin{eqnarray}
\hat{\rm A}_1&=&\hat{J}^+_1 \hat{J}^+_2 +\hat{J}^-_1 \hat{J}^-_2, \quad \hat{\rm B}_1=i(\hat{J}^+_1 \hat{J}^+_2-\hat{J}^-_1 \hat{J}^-_2 ), \\
\hat{ \tilde{{\rm A}}}_1&=&\hat{J}^+_1 \hat{J}^-_2 +\hat{J}^-_1 \hat{J}^+_2, \quad \hat{ \tilde{{\rm B}}}_1=i(\hat{J}^+_1 \hat{J}^-_2-\hat{J}^-_1 \hat{J}^+_2 ), 
 \end{eqnarray} 
and calculate
\begin{widetext}
\begin{eqnarray}
 \langle(\Delta \hat{\rm A}_1)^2\rangle_{_{\rm PT}}&=&\langle \hat{J}^+_1 \hat{J}^-_1 \hat{J}^+_2 \hat{J}^-_2+ \hat{J}^-_1 \hat{J}^+_1 \hat{J}^-_2 \hat{J}^+_2 \rangle + \langle \hat{J}^{+^2}_1  \hat{J}^{-^2}_2+ \hat{J}^{-^2}_1  \hat{J}^{+^2}_2\rangle - (\langle \hat{J}^{+}_1  \hat{J}^{-}_2+ \hat{J}^{-}_1  \hat{J}^{+}_2\rangle)^2 \nonumber  \\
 &=&\langle(\Delta \hat{ \tilde{{\rm A}}}_1)^2\rangle +4 \langle  \hat{J}^{(z)}_1  \hat{J}^{(z)}_2  \rangle, \\
 \langle(\Delta \hat{\rm B}_1)^2\rangle_{_{\rm PT}}&=&\langle \hat{J}^+_1 \hat{J}^-_1 \hat{J}^+_2 \hat{J}^-_2+ \hat{J}^-_1 \hat{J}^+_1 \hat{J}^-_2 \hat{J}^+_2 \rangle- \langle \hat{J}^{+^2}_1  \hat{J}^{-^2}_2+ \hat{J}^{-^2}_1  \hat{J}^{+^2}_2\rangle + (\langle \hat{J}^{+}_1  \hat{J}^{-}_2- \hat{J}^{-}_1  \hat{J}^{+}_2\rangle)^2 \nonumber \\
 &=&\langle(\Delta \hat{ \tilde{{\rm B}}}_1)^2\rangle +4 \langle  \hat{J}^{(z)}_1  \hat{J}^{(z)}_2  \rangle,\\
 \langle(\Delta \hat{\rm A}_1)^2\rangle_{_{\rm PT}}+ \langle(\Delta \hat{\rm B}_1)^2\rangle_{_{\rm PT}}&=& 4[\langle \hat{J}^+_1 \hat{J}^-_1 \hat{J}^+_2 \hat{J}^-_2\rangle+\langle  \hat{J}^{(z)}_1  \hat{J}^{(z)}_2  \rangle - |\langle \hat{J}^+_1 \hat{J}^-_2 \rangle|^2 ]-2 \langle\hat{J}^{(z)}_1  \hat{J}^+_2 \hat{J}^-_2+\hat{J}^{(z)}_2  \hat{J}^+_1 \hat{J}^-_1 \rangle \\ 
\langle[\hat{\rm A}_1,\hat{\rm B}_1]\rangle_{_{\rm PT}}&=&-4i\langle \hat{J}^+_1 \hat{J}^-_1 \hat{J}^{(z)}_2 + \hat{J}^{(z)}_1 \hat{J}^-_2 \hat{J}^+_2 \rangle,  \\
\langle \Delta \hat{\rm A}_1 \Delta \hat{\rm B}_1 \rangle_{_{\rm S, PT}} &=&  i \langle \hat{J}^{+^2}_1  \hat{J}^{-^2}_2- \hat{J}^{-^2}_1  \hat{J}^{+^2}_2\rangle - i (\langle \hat{J}^{+}_1  \hat{J}^{-}_2\rangle^2- \langle \hat{J}^{-}_1  \hat{J}^{+}_2\rangle^2), 
 \end{eqnarray}
 \end{widetext}
one can obtain the inequality in Eq.~(\ref{Eq-SR}). The elements of the above equations can also be written in terms of total angular momentum operator~($ \hat{J}_k $) and its components as  
\begin{widetext}
\begin{eqnarray}
\hat{J}^+_k \hat{J}^-_k&=&\hat{J}^{(x)^2}_k+\hat{J}^{(y)^2}_k+\hat{J}^{(z)}_k, \qquad  \hat{J}^-_k \hat{J}^+_k= \hat{J}^{(x)^2}_k+\hat{J}^{(y)^2}_k-\hat{J}^{(z)}_k, \nonumber \\
\langle \hat{J}^+_1 \hat{J}^-_1 \hat{J}^+_2 \hat{J}^-_2+ \hat{J}^-_1 \hat{J}^+_1 \hat{J}^-_2 \hat{J}^+_2 \rangle  &=&\langle \hat{J}^2_1 (\hat{J}^2_2- 2 \hat{J}^{(z)^2}_2)+ \hat{J}^2_2 (\hat{J}^2_1- 2 \hat{J}^{(z)^2}_1) +2 \hat{J}^{(z)}_1 \hat{J}^{(z)}_2 (\hat{J}^{(z)}_1 \hat{J}^{(z)}_2+1) \rangle,  \nonumber \\
\langle \hat{J}^{+^2}_1  \hat{J}^{-^2}_2+ \hat{J}^{-^2}_1  \hat{J}^{+^2}_2\rangle  &=& 2 ( \hat{J}^{(x)^2}_1-\hat{J}^{(y)^2}_1) (\hat{J}^{(x)^2}_2-\hat{J}^{(y)^2}_2) +2 \{ \hat{J}^{(x)}_1, \hat{J}^{(y)}_1 \}\{ \hat{J}^{(x)}_2, \hat{J}^{(y)}_2 \},  \nonumber  \\
(\langle \hat{J}^{+}_1  \hat{J}^{-}_2+ \hat{J}^{-}_1  \hat{J}^{+}_2\rangle)^2 &=& 4\langle( \hat{J}^{(x)}_1\hat{J}^{(x)}_2+\hat{J}^{(y)}_1\hat{J}^{(y)}_2)\rangle^2,\label{components} 
    \end{eqnarray}
\end{widetext}
where we use the relation: $ \hat{J}^\pm_k= \hat{J}^{(x)}_k \pm \hat{J}^{(y)}_k $ with k = 1, 2, and $ \{ \hat{J}^{(x)}_k, \hat{J}^{(y)}_k \} = \hat{J}^{(x)}_k \hat{J}^{(y)}_k+\hat{J}^{(y)}_k \hat{J}^{(x)}_k$. Similarly, to obtain spin squeezing-like criterion~[Eq.~(\ref{Eq-SR-ss})], one can use observables:
\begin{eqnarray}
\hat{\rm A}_2=\hat{J}^{(x)}_1  +\hat{J}^{(x)}_2,\qquad \hat{\rm B}_2=\hat{J}^{(y)}_1  +\hat{J}^{(y)}_2,
 \end{eqnarray}
and calculate
\begin{widetext}
\begin{eqnarray}
 \langle(\Delta \hat{\rm A}_2)^2\rangle_{_{\rm PT}}&=& \langle(\Delta \hat{\rm J}_1^{(x)})^2\rangle+\langle(\Delta \hat{\rm J}_2^{(x)})^2\rangle,   \\
  \langle(\Delta \hat{\rm B}_2)^2\rangle_{_{\rm PT}}&=& \langle(\Delta \hat{\rm J}_1^{(y)})^2\rangle+\langle(\Delta \hat{\rm J}_2^{(y)})^2\rangle,   \\
\langle[\hat{\rm A}_2,\hat{\rm B}_2]\rangle_{_{\rm PT}}&=& i\langle \hat{\rm J}_1^{(z)}-\hat{\rm J}_2^{(z)} \rangle, \\
\langle \Delta \hat{\rm A}_2 \Delta \hat{\rm B}_2 \rangle_{_{\rm S, PT}} &=& \langle  \hat{\rm J}_1^{(x)} \hat{\rm J}_1^{(y)} + \hat{\rm J}_1^{(y)} \hat{\rm J}_1^{(x)}- \hat{\rm J}_2^{(x)} \hat{\rm J}_2^{(y)} -\hat{\rm J}_2^{(y)} \hat{\rm J}_2^{(x)}\rangle   \nonumber \\ &-& 2 \langle \hat{J}^{(x)}_1 \hat{J}^{(y)}_2 + \hat{J}^{(y)}_1 \hat{J}^{(x)}_2\rangle   - 2 \langle \hat{J}^{(x)}_1  +\hat{J}^{(x)}_2 \rangle \langle \hat{J}^{(y)}_1  +\hat{J}^{(y)}_2 \rangle, 
 \end{eqnarray}
 \end{widetext}
to observe the results obtained in FIG.~\ref{fig-criteria}.

\subsection{Derivation of Eqs. (\ref{Eq:Jx})-(\ref{Eq:Jy})}
Here we show derivations of Eqs.(\ref{Eq:Jx})-(\ref{Eq:Jy}). To be short, we give the details of the calculations of the collective spin operators of the one ensemble. If one insert Eq.~(\ref{Displacement}) into Eq.~(\ref{Hols_Prim}) can get:
\begin{eqnarray}
\hat{J}_{1}^{+}&=&(\hat{d}_{1}^{\dagger} +\sqrt{{\rm N}_1 \alpha_1}) \sqrt{k} \nonumber \\
&\ast &\sqrt{1-\frac{\hat{d}_{1}^{\dagger}\hat{d}_{1}+(\hat{d}_{1}^{\dagger}+\hat{d}_{1})  \sqrt{{\rm N}_1 \alpha_1}}{k}},\\
& =& (\hat{d}_{1}^{\dagger} +\sqrt{{\rm N}_1 \alpha_1}) \sqrt{k} \sqrt{1-\xi}, \label{Eq:xi}
\end{eqnarray}
where $ k={\rm N}_1(1-\alpha_1) $.  After expanding the last term, $ \sqrt{1-\xi} $, in Eq.~(\ref{Eq:xi}) in the thermodynamic limit, one can arrive:
\begin{eqnarray}
\hat{J}_{1}^{+}&\approx & (\hat{d}_{1}^{\dagger} +\sqrt{{\rm N}_1 \alpha_1}) \sqrt{k}-\frac{{\rm N}_1 \alpha_1}{2\sqrt{k}} (\hat{d}_{1}^{\dagger}+\hat{d}_{1}),
\end{eqnarray}
where we obtain powers of $ {\rm N}_1 $ up to zeroth order. Similarly, one can derive lowering component, $\hat{J}_{1}^{-}  $. By using definition: $ \hat{J}_{1}^{(x)}=({\hat{J}_{1}^{+}+\hat{J}_{1}^{-}})/{2} $ and $ \hat{J}_{1}^{(y)}=({\hat{J}_{1}^{+}-\hat{J}_{1}^{-}})/{2i} $ and Eqs. (\ref{Eq-posx})-(\ref{Eq-posy}), we derive Eqs. (\ref{Eq:Jx})-(\ref{Eq:Jy}) as
\begin{eqnarray}
\hat{J}_1^{(x)}\cong& &  {\rm N}_1 \sqrt{ \alpha_1 (1-\alpha_1)}+\sqrt{\frac{{\rm N}_1 \Omega_1}{2}} \frac{1-2\alpha_1}{\sqrt{1-\alpha_1}}\hat{X}, \\
\hat{J}_1^{(y)} \cong &&-\sqrt{\frac{{\rm N}_1 (1-\alpha_1) }{2 \Omega_1}}\hat{P}_{X}.
\end{eqnarray}

\bibliography{bibliography}

\end{document}